\documentstyle[aps,preprint,epsfig]{revtex}

\def\Vec#1{\mbox{\boldmath $#1$}}

\begin{document}
\title{A realistic study of  the nuclear transparency \\
 and the distorted momentum distributions\\ in the semi-inclusive process $^4He(e,e'p)X$ }
\draft

\author{H. Morita\thanks{On leave of absence  from Sapporo Gakuin University, Bunkyo-dai 11, Ebetsu 069, Hokkaido, Japan} and C. Ciofi degli Atti }
\address{Department of Physics, University of Perugia, and Istituto Nazionale di Fisica Nucleare, Sezione di Perugia, Via A. Pascoli, I--06100 Perugia, Italy}
\author{D. Treleani}
\address{Department of Theoretical Physics, University of Trieste, Strada
Costiera 11, Istituto Nazionale di Fisica Nucleare, Sezione di Trieste, and ICTP, I--34014, Trieste,Italy}
\date{\today}
\maketitle

\begin{abstract}
The nuclear transparency and the distorted momentum distributions of $^4$He in the semi-inclusive process $^4He(e,e'p)X$ are calculated within the Glauber multiple scattering approach using for the first time realistic four-body variational wave functions embodying central and non-central nucleon-nucleon (N-N) correlations. The contributions from N-N correlations and from Glauber multiple scattering are taken into account exactly to all orders. It is shown that non-central correlations significantly affect both the transparency and the distorted momentum distributions; as a matter of fact: i) the small ($\approx 3\%$) value of the effect of correlations on the transparency results from an appreciable cancellation between the short-range central repulsive correlations and the intermediate-range  attractive correlations, whose magnitude is significantly affected by the non-central forces, and ii) the effect of Glauber final state interactions on the momentum distribution is
reduced by the inclusion of tensor  correlations. 
\end{abstract}

\section{Introduction}

The role played by ground state nucleon--nucleon(NN) correlations (or {\it initial state correlations} (ISC))  on the nuclear transparency in semi-inclusive  processes have been   investigated by many authors. (see e.g. \cite {ref:nikolaev}- \cite {ref:strikman}) with conflicting results. All of these works adopt the Glauber multiple scattering approach for the description of the final state interaction (FSI), and treat ISC within different schemes and approximations, ranging from phenomenological Jastrow wave functions embodying only central correlations \cite {ref:bianconi}, \cite {ref:bianco} to various expansions in terms of correlated density matrices  \cite {ref:benar}, \cite {ref:rinat}, \cite {ref:kohama}. Recently, Seki et al.\cite{ref:seki} have performed an elaborated calculation of the transparency using, by means of a  local density approximation (LDA),  realistic nuclear matter correlation functions for  finite nuclei, and taking into account Glauber multiple scattering to all orders. Such a calculation   represents a significant progress in the field, but  it should  be pointed out that a direct calculation of the transparency using realistic wave functions resulting from the solution of the many-body problem and implemented by Glauber multiple scattering operators, is still lacking. It is the purpose of this paper to present such a calculation for the process $^4He(e,e'p)X$, for which we have calculated both the nuclear transparency  and the distorted momentum distributions $n_D(k)$. The latter quantity has been recently calculated in Ref. \cite{ref:bianco}, where it has been argued  that the high momentum part of  $n_D(k)$  which could be measured by semi-inclusive processes,
is almost entirely dominated by FSI, leaving little room for the investigation of ISC. The results of \cite{ref:bianco} are based on the use of phenomenological Jastrow wave functions constructed from harmonic oscillator orbitals and simple central correlation functions.
Since realistic nuclear wave functions exhibit a complex correlation structure 
generated  by the very nature of the NN interaction, which cannot be recoinciled with the Jastrow correlation function, it is imperative to consider the effect of FSI within a realistic treatement of ISC. This is precisely the aim of the present paper where, as in Ref.\cite{ref:bianco}, Glauber multiple scattering is taken into account exactly to all order, but, at the same time, wave functions embodying realistic central, tensor, spin, isospin, etc. correlations are. for the firdt time, employed in this kind of calculations. Thus, our approach is basically the same as the one of Ref. \cite{ref:bianco} (i.e. Glauber multiple scattering and ISC taken into account to all orders), with the substantial difference of using a realistic four body wave function.

Our paper is organised as follows: in Section \ref{sec:crosssection} the concepts of semi-inclusive processes, distorted momentum distributions and nuclear transparency,  will be racalled;the structure of the $^4$He wave function used in the calculations will be briefly described in Section \ref{sec:atms}; the results of calculations of the nuclear transparency and the distorted momentum distributions will be presented in Sections \ref{sec:transparency} and \ref{sec:momentum}, respectively; the Summary and Conclusions are given in Section \ref{sec:conclusion}.

\section{The cross section for the semi-inclusive {\bf\it A(\lowercase{e,e'p})X} 
process, the distorted  momentum distributions and the nuclear transparency}
\label{sec:crosssection}
In this Section  we will consider the process $A(e,e'p)X$ in which an electron with 4-momentum
$k_1\equiv\{{\bf k}_1,i\epsilon_1\}$, is scattered off a nucleus with 4-momentum
$P_A\equiv\{{\bf 0},iM_A\}$ to a state $k_2\equiv\{{\bf k}_2,i\epsilon_2\}$ 
and is detected in coincidence with a proton $p$ with 4-momentum 
$k_p\equiv\{{\bf k}_p,iE_p\}$; the final $(A-1)$ nuclear system with 
4-momentum $P_X\equiv\{{\bf P}_X,iE_X\}$ is undetected. From now on, we will closely follow the formalism of  Ref. \cite {ref:nikolaev},
\cite {ref:bianconi}-\cite{ref:bianconi2}, and, accordingly,  write the 
coplanar
cross section 
describing the process in the following form

\begin{equation}
\frac{d\sigma}{dQ^2d\nu d{\bf k}_p}=K\sigma_{ep}P_D(E_m,{\bf k}_m,{\bf k}_p)
\label{sezione}
\end{equation}
where $P_D(E_m,{\bf k}_m,{\bf k}_p)$ is the distorted nucleon spectral function, $K$  a kinematical factor, $\sigma_{ep}$ the off-shell electron-nucleon 
cross section,  and $Q^2=|{\bf q}|^2-\nu^2$ the four momentum transfer, with   the $z$ axis is oriented along $\bf q$.
Eq. 1 has been obtained under the assumption that  the  difference between the longitudinal (L) and transverse (T) spectral functions arising from the absorption of longitudinal and transverse photons has been disregarded, which means that  spin effects in the FSI have been neglected due to   the large energy of the struck proton; this  means that the distorted spectral function will be  evaluated by using the electromagnetic charge operator only.  In Eq. 1 
\begin{equation}
{\bf k}_m={\bf q}-{\bf k}_p
\label{missingmom}
\end{equation}
is the {\it missing momentum}
and 
\begin{equation}
E_m=\nu+M-E_p - T_{X}^R.
\label{missingen}
\end{equation}
the {\it missing energy},
where $T_{X}^R$ is the kinetic recoil energy of $X$.
The latter equation results from energy conservation

\begin{equation}
\nu+M_A=E_p+\sqrt{M_X^2+{\bf p}_X^2}
\label{energycons}
\end{equation}
if the total energy of the system $X$ is
approximated by its non-relativistic
expression.
Following Ref. \cite {ref:nikolaev},
\cite {ref:bianconi}-\cite{ref:bianconi2}, the longitudinal distorted spectral function 
is  written in the following form
\begin{equation}
P_D(E_m,{\bf k}_m)=\sum_{f_X}|\langle{\bf 
k}_m,\Psi_{f_X}|\Psi_A\rangle|^2\delta(E_m-E_{min}-E_{f_X})
\label{Pempm}
\end{equation}
where $E_{min} = M+M_{A-1}-M_A$, and 

\begin{equation}
\langle{\bf k}_m,\Psi_{f_X}|\Psi_A\rangle=\int e^{i{\bf k}_m\cdot{\bf r}_A}S_G({\bf 
r}_1\dots{\bf r}_A)\Psi_{f_X}^*({\bf r}_1\dots{\bf r}_{A-1})\Psi_A({\bf 
r}_1\dots{\bf r}_A)\delta ({\sum_{j=1}^A {\bf r}_j})\prod_{i=1}^Ad{\bf r}_i,
\label{overlap}
\end{equation}
with $\Psi_A$ and $\Psi_{f_X}$ being the ground state wave 
function of the target nucleus and the wave function of the system $X$ in the 
state $f_X$, respectively; the quantity $S_G$ is the Glauber operator, which describes the  FSI of the struck proton with the $(A-1)$ system, i.e. 

\begin{equation}
S_G({\bf r}_1\dots{\bf r}_A)=\prod_{j=1}^{A-1}G({\bf r}_A,{\bf r}_j)\equiv
\prod_{j=1}^{A-1}\bigl[1-\theta(z_j-z_A)\Gamma({\bf b}_A-{\bf b}_j)\bigr]
\label{SG}
\end{equation}
where ${\bf b}_j$ and $z_j$ are the transverse and the longitudinal components 
of the nucleon coordinate ${\bf r}_j\equiv({\bf  b}_j,z_j)$, ${\mit\Gamma}({\bf 
b})$ is
the Glauber profile function for elastic proton nucleon scattering, and the 
function $\theta(z_j-z_A)$ takes care of the fact that the struck proton ``A'' 
propagates along a straight-path trajectory so that it interacts with nucleon 
``$j$'' only if $z_j>z_A$. The integral over the missing energy of the distorted 
spectral function defines the distorted momentum distribution as

\begin{equation}
n_D({\bf k}_m)=\int dE_m P_D(E_m,{\bf k}_m).
\label{nd}
\end{equation}

In impulse approximation (IA) (i.e. when the final state interaction is 
disregarded ($S_G=1$)), if the system $X$ is assumed to be a $(A-1)$ nucleus in 
the discrete or continuum states $f_X\equiv f_{A-1}$, the distorted spectral function 
$P_D$ reduces to the usual spectral function i.e.

\begin{equation} 
P_D\to P(k,E)=\sum_{f_{A-1}}|\langle{\bf 
k},\Psi_{f_{A-1}}|\Psi_A\rangle|^2\delta\bigl(E-(E_{min}+E_{f_{A-1}})\bigr
)
\label{Pke}
\end{equation}
where $E$ is the nucleon removal energy i.e. the energy required to remove a 
nucleon from the target, leaving the $A-1$ nucleus with excitation energy 
$E_{f_{A-1}}$ and ${\bf k}=-{\bf k}_m=-({\bf q}-{\bf k}_p)$ is the nucleon momentum 
before interaction. The integral of the spectral function over the $E$ defines the (undistorted)
momentum distributions 
\begin{equation}
n({\bf k})=\int dE P(E,{\bf k}).
\label{nk}
\end{equation}

In this paper we will consider the effect of the FSI ($S_G\not=1$) on the 
semi-inclusive $A(e,e'p)X$ process, i.e.  the cross section (\ref{sezione})
integrated over the missing energy $E_m$, at fixed value of $\bf k_m$. Owing to 

\begin{equation}
\sum_{f_X}\Psi_{f_X}^*({\bf r}_1'\dots{\bf r}_{A-1}')\Psi_{f_X}({\bf r}_1\dots{\bf 
r}_{A-1})=\prod_{j=1}^{A-1}\delta({\bf r}_j-{\bf r}_j')
\label{closure}
\end{equation}
the cross section (\ref{sezione}) becomes directly proportional to the distorted 
momentum distributions (\ref{nd}), i.e. 

\begin{equation}
n_D({\bf k}_m)={(2 \pi)^{-3}} \int e^{i {\bf k}_m\cdot({\bf r} -{\bf r}')}\rho_D 
({\bf r},{\bf r}') d{\bf r} d{\bf r}'
  \label{nd1}
   \end{equation}
where 
\begin{eqnarray}
\rho_D ({\bf r},{\bf r}')= \frac {\langle\Psi_A S_G^{\dagger} \hat{O}({\bf 
r},{\bf r}')  S_G'\Psi_A'\rangle}{\langle\Psi_A\Psi_A\rangle}
  \label{rodi}
   \end{eqnarray}
is the  one-body mixed density matrix, and
\begin{eqnarray}
\hat{O}({\bf r},{\bf r}')= \sum_i\delta({\bf r_i} - {\bf r}) \delta(\bf r_i^{'} - \bf r^{'})\prod_{j\not= i}\delta({\bf r_j}-{\bf r_j}')
\label{rodiop}
   \end{eqnarray}
the one-body density operator.
In Eq.~(\ref{rodi}) and in the rest of the paper, the primed quantities have to 
be evaluated at ${\bf r}'$. By integrating $n_D({\bf k}_m)$
one obtains the nuclear transparency $T$, which is defined as follows

\begin{eqnarray}
T=\int n_D({\bf k}_m) d{\bf k}_m = (2 \pi)^{-3}\int \rho_D ({\bf r},{\bf r}') d{\bf r} d{\bf 
r}'
\int e^{i {\bf k}_m\cdot({\bf r} -{\bf r}')}d{\bf k}_m = \int \rho_D ({\bf r})d {\bf 
r}
  \label{intnd}
   \end{eqnarray}
i.e.

\begin{eqnarray}
T = \int\rho_D ({\bf r})d {\bf r} = 1+ \Delta T
  \label{ti}
   \end{eqnarray}
where $\Delta T$ originates from FSI.
   
We specialize now to the four-body system, for which the use of intrinsic coordinates is mandatory. The one-body mixed density matrix then becomes

\begin{eqnarray}
  \rho_D ({\bf r},{\bf r}')= \int d{\bf R}_1d{\bf R}_2 \Psi^{*}({\bf R}_1,{\bf R}_2,{\bf R}_3={\bf r})\ \hat{S_G}^{\dagger}\,\hat{S_G}'\ \Psi({\bf R}_1,{\bf R}_2,{\bf R}_3'={\bf r}'),
  \label{eq:rhod}
\end{eqnarray}
where $\Psi$ is the four-body intrinsic nuclear wave function  normalized to unity  and expressed in terms of the following Jacobi coordinates 
\begin{eqnarray}
  {\bf R}_1&=&{\bf r}_2-{\bf r}_1,\nonumber \\
  {\bf R}_2&=&{\bf r}_3-({\bf r}_2+{\bf r}_1)/2,\nonumber \\
  {\bf R}_3&=&{\bf r}_4-({\bf r}_3+{\bf r}_2+{\bf r}_1)/3.
  \label{eq:jacobi}
\end{eqnarray}
The operator $\hat{S_G}$ given by Eq.~(\ref{SG}) is now explicitly written as 
\begin{eqnarray}
  \hat{S_G}=\prod_{i=1}^3 G(4i),  \qquad G(4i)=1- \theta(z_4-z_i)\Gamma (\Vec{b}_i-\Vec{b}_4),
  \label{eq:s}
\end{eqnarray}
where the longitudinal component $z_{4}$ is   along the direction of the momentum transfer, which
is chosen to coincide with the z-axis, thus ${\bf r}_4$ is expressed as 
${\bf r}_{4}= {\bf b}_{4}+z_{4}\hat{{\bf q}}$.
The standard parametrization for the  profile function $\Gamma$  viz. 
\begin{eqnarray}
  \Gamma(\Vec{b})= \frac{\sigma_{tot}(1-i\alpha)}{4\pi b_o^2}e^{-\Vec{b}^2/2b_0^2}  \label{eq:gamma}
\end{eqnarray}
will be adopted, 
where $\sigma_{tot}$ is the total proton-nucleon cross section, and $\alpha$ the ratio of the real to the imaginary parts of the forward elastic pN scattering amplitude. In the calculations the values  $\alpha=-0.33$,  $b_0=0.5 fm$ and $\sigma_{tot}=43 mb$ have been used,  which correspond to a kinetic energy of the hit proton $T_p \approx 1GeV$.

The distorted momentum distribution $n_D({\bf k}_m)$ and the nuclear transparency $T$ are calculated from Eqs.~(\ref{nd1}) and (\ref{intnd}) respectively with $\rho_D$ of Eq.~(\ref{eq:rhod}).
%

\section{The realistic four-body  wave function}\label{sec:atms}
The $^4$He wave function used in our calculations has been obtained by the
variational  ATMS method\cite{ref:atms1,ref:atms2}, according to which 
\begin{equation}
\Psi_{ATMS} = F\cdot\Phi,
\label{eq:atmswf}
\end{equation}
 where $F$ represents a proper correlation function, and  $\Phi$ is an arbitrary uncorrelated wave function. The correlation function $F$ has the following form 

\begin{mathletters}
\begin{eqnarray}
\label{eq:atms_cor}
F&=&D^{-1}\sum_{ij}(w(ij)-\frac{(n_p-1)}{n_p}u(ij))\prod_{kl \neq ij}u(kl),\\
D&=&\sum_{ij}(1-\frac{(n_p-1)}{n_p}u(ij))\prod_{kl \neq ij}u(kl),
\end{eqnarray}
\end{mathletters}
where $n_p=A(A-1)/2$ is the number of pair, and $w(ij)$ ( $u(ij)$ ) are on-shell ( off-shell ) two-body correlation functions, respectively.
The wave function $\Phi$ is assumed to be 

\begin{equation}
\Phi = \Phi_S \cdot \{S=0,T=0\}_A,
\label{eq:atms_init}
\end{equation}
where $\{0,0\}_A$ represents the antisymmetric spin-isospin function \cite{ref:ST} with $S=T=0$, while $\Phi_S$ is the fully symmetric spatial function. 

The realistic NN interaction generates a state dependence of NN correlations, which is taken into account by introducing the following state dependence for the on-shell correlation function

\begin{equation}
w(ij)=^1w_S(ij)\hat{P}^{1E}(ij) + ^3w_S(ij)\hat{P}^{3E}(ij) + ^3w_D(ij)\hat{S}_{ij}\hat{P}^{3E}(ij),
\label{eq:atms_onshell}
\end{equation}
where $\hat{P}^{1E}$($\hat{P}^{3E}$) is a projection operator to the singlet-even (triplet-even) state and $\hat{S}_{ij}$ is the usual tensor operator. Tensor-type off-shell correlations are also included in $F$  (the explicit form of $\Psi_{ATMS}$ is given in \cite{ref:atms3}). The best set of correlation functions $\{u\}=\{w's,u's\}$ appearing in Eq.~(\ref{eq:atmswf}) are determined by the Euler-Lagrange equation

\begin{equation}
\delta_{\mu}[<\Psi_{ATMS}|H|\Psi_{ATMS}>-E<\Psi_{ATMS}|\Psi_{ATMS}>] = 0,
\label{eq:atms_euler}
\end{equation}
where the variation $\delta_{\mu}$ is performed with respect to each correlation function $u_{\mu}$.
If a pair product form is assumed for $\Phi_S$ 

\begin{equation}
\Phi_S = \prod_{ij} \phi(ij),
\label{eq:atms_phi}
\end{equation}
then the radial shape of the two-body function $f^{\mu}_{ATMS}(ij)=u_{\mu}(ij)\phi(ij)$ can be detrmined directly from Eq.~(\ref{eq:atms_euler}). Thus the ATMS wave function does not contain any free parameter. In the rest of the paper, we use for the realsitic NN interaction, the Reid soft core $V8$ model potential.\cite{ref:reidv8}
The calculated binding energy corresponding to the ATMS wave functions is $E_4 =-21.2 MeV$, the $rms$ radius is $<r^2>^{1/2} = 1.57 fm$, and the probabilities of the various waves are $P_{S}=87.94 \%$, $P_{S'}=0.24 \%$ and $P_{D}= 11.82 \%$.

\section{Results of the calculations: the nuclear transparency }\label{sec:transparency}

In this section the results of the calculations of the nuclear transparency will be presented. 

Let us  first discuss the much debated topic concerning the effects of NN correlations on the transparency. To this end, as a $\Phi_S$ of Eq.~(\ref{eq:atms_phi}) we introduce a mean-field wave function $\phi_0$  defined by an harmonic oscillator state  with the correct tail\cite{ref:tan}, viz. 

\begin{mathletters}
\begin{eqnarray}
  \phi_{0} &=& \prod_{ij} \phi(r_{ij}),\\
  \phi(r) &=& N
    \left\{
      \begin{array}{cr}
        \ e^{-r^2/8R_0^2}            & \qquad r < r_0 \\
        N'\frac{e^{-\beta r/3}}{r^{1/3}}   & \qquad r \ge r_0 ,
      \end{array}
    \right.
  \label{eq:phi0}   
\end{eqnarray}
\end{mathletters}
where the quantities $r_0$ and $N'$ are determined by the continuity condition, and $\beta$ and $R_0$ are fixed so as to reproduce, respectively, the asymptotic behaviour of the realistic two-body function $f_{ATMS}(r)$ and the root mean square radius ($1.57 fm$) provided by the realistic wave function, i.e. $\beta=0.76$  and  $R_0=1.22 fm$. Thus, by calculating the nuclear transparency using the full realistic, $\Psi_{ATMS}$,  and the mean field, $\phi_0$, wave functions, one can ascribe any difference in the results to the effect of the correlations.

The results of the calculation are presented in Table~\ref{tab:transparency}, where it can be seen that the effect of correlations on the transparency,
given by  the difference between $T_{ATMS}$ and $T_0$,  is very small ($\sim 3\%$). 
We have also calculated $T$ with the Jastrow-type wave function 

\begin{eqnarray}
  \Psi_{J} &=& \prod_{ij} f_J(r_{ij})\phi_0, \qquad  f_J(r_{ij}) = 1- e^{-r_{ij}^2/2r_c^2}, 
  \label{eq:jastrow}
\end{eqnarray}
used in \cite{ref:bianconi} with $r_c=0.5 fm$. It can be seen from Table~\ref{tab:transparency} that in such a case correlation effects amount to $\sim 7\%$, which is about a factor of two larger than in the realistic case. Since the Jastrow-type correlation function in Eq.~(\ref{eq:jastrow}) takes only into account short-range repulsive correlations, the difference between the realistic and the Jastrow cases should be ascribed to the intermediate-range attractive correlation which is provided by realistic correlation function. In order to clearly illustrate this point, the realistic correlation function $^3\tilde{w}_s(r)=\ ^3w_s(r)-(5/6)u(r)$ is compared with the Jastrow one in Fig.~\ref{fig:correlation}, and the strong overshooting in the  function, which is induced by the attractive correlation, and which is lacking in the Jastrow phenomenological wave function, can be noticed. 
The attractive correlation enhances FSI effects and, as a result, the nuclear transparency is reduced because of the cancellation between short-range repulsive and intermediate-range attractive correlations.

The tensor-type correlations, which are induced by the realistic interaction,  produce  a D-wave component in the $^4$He wave function. The value of the nuclear transparency calculated without the D-wave component,  denoted $T_S$ in table~\ref{tab:transparency},  does not appreciably differ from the full results ($T_{ATMS}$). However, one should not share the mistaken believe that tensor correlations are unimportant, because the shape of the correlation function shown in Fig.~\ref{fig:correlation}, which determines the relative strength between repulsive and attractive correlations, is strongly affected by the tensor force through the S-D wave coupling. Therefore, the inclusion of tensor correlations is essential in determining the exact correlation function with the proper overshooting, even if the presence of 
the D-wave in the wave functions provides only a small contribution to the transparency.

It has been argued \cite{ref:nikolaev} that the effect of correlations on the transparency 
should be very small, because of the  cancellation between the {\it hole} and {\it spectator} effects. In order to check whether such a conclusion holds in  a realistic approach to correlations,  we have calculated the {\it hole} contribution to the transparency, by removing correlations among all the spectator pairs   setting $f_{ATMS}(ij)\rightarrow\phi_0(ij)$ for all spectator-(ij) pairs. The result, denoted by $T_{ATMS}($no spect.$)$ in table~\ref{tab:transparency},
practically does not differ from the full reuslt $T_{ATMS}$, which includes {\it hole} and {\it spectator} correlations. This clearly shows that the {\it spectator} effect is extremely small and that there is no significant cancellation between {\it hole} and {\it spectator} correlations.

Let us now discuss the convergence of the Glauber multiple scattering series. To this end, using Eq.~(\ref{eq:s}), the operator $\hat{S_G}^{\dagger}\hat{S_G}$ in Eq.~(\ref{eq:rhod}) is expanded in the following way  

\begin{mathletters}
\begin{eqnarray}
  \hat{S_G}^{\dagger}\hat{S_G} &=& 1 + \Delta G_1 + \Delta G_2 + \Delta G_3 , \\
  \Delta G_1 &=& \sum_{i<4} (|G(4i)|^2-1) , \nonumber \\
  \Delta G_2 &=& \sum_{i<j<4} (|G(4i)|^2-1)(|G(4j)|^2-1), \\
  \Delta G_3 &=& \prod_{i<4} (|G(4i)|^2-1). \nonumber
  \label{eq:g_expansion}
\end{eqnarray}
\end{mathletters}
Several approaches in the field (\cite{ref:rinat}, \cite{ref:kohama}) rely on the truncation of the Glauber multiple scattering series taking into account only the single rescattering term $\Delta G_1$.  In table \ref{tab:Glauber} the contribution to the transparency from the various terms of the series in Eq.~(29) are presented. It can be seen that $\Delta G_2$ amounts to $14\%$ of the first order contribution $\Delta G_1$, and its inclusion changes the total $T$ by about $5\%$. Therefore the double rescattreing term $\Delta G_2$ should not be disregarded in the calculation of the transparency.

\section{Results of the calculations: the distorted momentum distributions}\label{sec:momentum}
In this section the results of the calculations of the distorted momentum distributions will be presented.  We will compare our results with the ones obtained in \cite{ref:bianconi} using phenomenological Jastrow wave functions adopting the same form and parameters for $\Psi_{J}$ in E
q.~(\ref{eq:jastrow}), namely the harmonic oscillator wave function $\phi_0$ with parameter $R_0=1.29fm^{-1}$, and the central correlation function with parameter $r_c=0.5fm^{-1}$. We will consider the {\it parallel} 
$(\theta =0^o, k_{\perp}= 0)$, {\it antiparallel}  $(\theta =180^o, k_{\perp}= 0)$,
and {\it perpendicular}$   (\theta =90^o, k_{\parallel}= 0)$ momentum distributions, where  $\theta$ is the angle between the three-momentum transfer ${\bf q}$ and ${\bf k_m}={\bf q}-{\bf k_p}$. From now on, the  notation $\bf k \equiv \bf k_m$  will be used.  
In absence of FSI the three distributions will coincide with the usual momentum distribution $n(\mathbf{k})$. The realistic momentum distributions $n(\mathbf{k})$ corresponding to the ATMS method are shown in Fig.~\ref{fig:ennek}, where they are compared with the momentum distributions corresponding to the Jastrow wave function. The numerical integration  of $n(\mathbf{k})$ with realistic wave functions has been performed by the quasi random number ($QRN$) method \cite{ref:QRN}.

Let us consider the distorted momentum distributions and let us demonstrate that the relevance of FSI effects strongly depends upon the  type of ground state wave functions. To this end, we  consider the mean field result given by the harmonic oscillator model wave function and the realistic one calculated  by the ATMS wave function. The results for the parallel distorted momentum distributions are shown in Fig.~\ref{fig:gr+FSI}. It can be seen that:
\begin{enumerate}
\item {as expected, the main effect of  FSI is to enhance the content of high momentum components (cf. Fig.~3(a));}
\item {FSI are reduced when the correct high momentum content of realistic ground state wave functions is taken into account (cf. Fig.~3(b))}.
\end{enumerate}

Let us now compare the effects of FSI in the realistic and Jastrow cases. To this end, the   undistorted and  distorted (anti)parallel and perpendicular  momentum distributions are compared in Fig.~\ref{fig:mdis_overview}. The results there shown,  clearly exhibit  again a strong dependence  of FSI   upon the  type of
ground state  correlations; it can in particular be seen that FSI are much larger in the Jastrow case. Only in the case of perpendicular kinematics, the distorted momentum distributions appear to be the same in the Jastrow and realistic cases: at $\theta = 90^{\circ}$, Glauber multiple scattering FSI produces the largest content of high momentum components. Eventually, in Fig.~\ref{fig:comparison} the overall comparison between the realistic and the Jastrow distorted momentum distributions is presented and the following points are worth being noticed:
\begin{enumerate}
	\item { a sizeable difference between Jastrow and realistic wave functions shows up in the low momentum region; such a difference 
should be ascribed  both to the tail behaviour of the wave functions, with 
the Jastrow function having the wrong asymptotics, and  to the attractive correlation effect discussed in the previous Section.}
\item { the longitudinal momentum distributions appreciably differ in the high momentum region $k \ge 2.0 fm^{-1}$ , which is mainly due to the effect of the D-wave, which is absent in the Jastrow wave function. The effect of the D-wave component will be discussed below in more details.}
\end{enumerate}

In order to better visualize the effect of FSI, the quantity $R(\Vec{k})=n_D(\Vec{k})/n(\Vec{k})$, i.e. the ratio between the distorted and undistorted momentum distributions,  is presented in Fig.~\ref{fig:ratio}. It can be seen that while the parallel and anti-parallel ratios are of the order of unity, the perpendicular ratio may reach (in the Jastrow case) almost an order  of magnitude. This is a natural result which comes from the Glauber profile function $\Gamma$ in Eq.~(\ref{eq:gamma}); in fact, due to its short-range nature in the $\Vec{b}$ plane, it creates a high momentum components at transverse directions. Thus at perpendicular kinematics, FSI  dominates the high momentum component induced by ground state correlations, as pointed out in ref.~\cite{ref:bianconi}. However, from a quantitative point of view, its magnitude largely depends upon the nature of correlations, as clearly illustrated by the differences exhibited in Fig.~\ref{fig:ratio} by Jastrow and realistic wave functions. The smaller effect of FSI in the realistic case is mainly due to the D-wave component produced by the tensor-type correlations. As a matter of fact, the D-wave component, being more peripheral with respect to the S-wave component, is less affected by FSI, since for a peripheral nucleon the probability of a collision is reduced. This is clearly 
demonstrated  in Fig.~\ref{fig:angular} where the angular dependence of the distorted momentum distribution is plotted for a fixed value of the missing momentum $k$. The figure shows indeed that the D-wave component (dot-dashed curve) is little affected by FSI (remember that the undistorted momentum distribution is angle independent).
The effects of the $D$ wave in the deuteron in the process $^2H(e,e'p)X$, has been investigated in Ref. \cite {ref:bianconi2}; the results obtained here for $^4He$ are qualitatively similar to the ones found there, with expected and obvious quantitative differences, due to the different role played by the $D$ wave in the four-body system.
An intersting quantity is the forward-backward asymmetry $A_{FB}$
 \begin{eqnarray}
 A_{FB}(k) = \frac{n_D(k:\theta=0^{\circ})-n_D(k:\theta=180^{\circ})}{n_D(k:\theta=0^{\circ})+n_D(k:\theta=180^{\circ})} .
 \label{eq:afb}
 \end{eqnarray}
which has been first introduced in  \cite{ref:bianconi2} in the analysis of the process
$^2H(e,e'p)X$.
The interest in the asymmetry, which differs from zero because of the FSI, stems from the following reason: since, as previously illustrated  in Fig.~\ref{fig:angular}, the D-wave component has little asymmetry, $A_{FB}$ gets contribution almost entirely from the $S$-wave. This suggests that this quantity might give an useful information on the S-wave component. To demonstrate such a point, 
we have calculated  $A_{FB}$ with the realistic wave function by changing the D-wave probability $P_D$ within a reasonable range \cite {ref:rosati}. The results are presented in
Fig.~\ref{fig:asymmetry}, which indeed shows that $A_{FB}$ slightly depend upon the $^4He$ $D$-wave probability. In the Figure the predictions by the Jastrow wave function are also shown (\cite{ref:bianco}) and, again, it can be seen that they are qualitatively similar to the realistic ones, but quantitatively different.
\section{Summary and Conclusions}\label{sec:conclusion}
 In this paper we have thoroughly investigated the ISC and FSI effects of correlations on the  nuclear transparency and distorted momentum distributions of of the semi-inclusive process $^4He(e,e'p)X$; FSI were treated within 
 the Glauber approach, according to \cite{ref:nikolaev}, whereas the effects of ISC result from the use of a variational  four-body  wave function, corresponding to a realistic nucleon-nucleon interactions producing central, tensor, spin and  isospin
 correlations. 
Our main results can be summarised as follows:
 \begin{enumerate}
 	\item The effect of realistic NN correlations on nuclear transparency $T$ amounts to $\sim 3\%$. Such a  small value mainly results from the  cancellation between short-range repusive correlations and intermediate-range attractive correlations originating from the central-tensor coupling, and not from the cancellation between the {\it hole} and the  {\it spectator}  correlation contributions, for the latter  is always much smaller than the former  and thus there is no significant cancellation between them (the same appears to hold   for complex nuclei as well  \cite{ref:ciofi1}, \cite{ref:cladahiko}, \cite {ref:ciofi2}).  The contribution from double rescattering term, e.g. the {\it ternary collision term} \cite{ref:rinat} amounts to $14\%$ of the leading order term, the  single rescattering contribution. It is clear therefore that the double rescattreing contribution should be taken into account even at high momentum transfer region, such as  $Q^2\approx$  several $(GeV/C)^2$, which is the kinematical region considered in this paper.
\item The effect of correlations and Glauber FSI on the distorted momentum distributions appreciably depends upon the type of wave functions which are adopted to describe the  ground state; if ground-state correlations are completely absent, a situation which does not occur in nature, for real nuclei are strongly correlated systems, the whole high momentum content of the distorted momentum distributions  is provided by FSI. 
However, if ISC are considered, the role of FSI is appreciably suppressed (cf. Fig. \ref{fig:gr+FSI}). We would like to stress, in this connection, that introducing FSI without, at the same time, considering ISC, is a pure academic and misleading operation, for, we reiterate it once again, realistic nuclear wave functions contains a large amount of ISC.
The issue we addressed in the present paper was whether and to which extent  the
central Jastrow
correlations could mock up the realistic ones in the semi-inclusive process $A(e,e'p)X$. According to our results, the answer is that, at high missing momenta,  the Jastrow and realistic approaches may differ by a factor of $\simeq {2-3}$,  in the parallel and antiparallel kinematics, whereas in perpendicular kinematics they do not appreciably differ (cf. Fig. \ref{fig:comparison}); moreover,  the Jastrow approach always overestimates FSI by the same  factor  (cf. Fig. \ref{fig:ratio}).
\item The effect of FSI dominates the high momentum component of the perpendicular distorted momentum distributions $n_D({\bf k}_{\perp}, k_{\parallel}=0)$, though its magnitude is reduced if  tensor-type correlations which induce the D-wave component in $^4$He are taken into account.
   \item  The parallel (antiparallel) distorted momentum distributions  $n_D({\bf k}_{\perp}={0},k_{\parallel})$ are appreciably affected at medium and high momenta by FSI if  uncorrelated or Jastrow centrally correlated wave functions are adopted to describe the $^4$He ground state; if, however, the latter is described by ''real" four-body wave functions, featuring the correct high momentum content, the effects of FSI on the high momentum part of $n_D({\bf k}_{\perp}={0},k_{\parallel})$ is, as already pointed out, 
reduced by a factor of $\simeq {2-3}$; nevertheless, a difference of about $20-30 \%$ still persists between distorted and undistorted momentum distributions; 
therefore,  FSI (as well as Meson Exchange Currents, Isobar configurations, etc.) have   to be
always carefully taken into account in the analysis of semi-inclusive processes,
in order to infer whether different momentum distributions produced by  different {\it realistic many body wave functions} can be discriminated by measuring the {\it distorted} momentum distributions.
\item The forward-backward asymmetry $A_{FB}(k)$  represents an interesting quantity even for the four-body system, due to its ability to filter out the S-wave component of the  wave function.
\end{enumerate}

Calculations of the distorted momentum distribution for complex nuclei, using  a  recently proposed approach to treat   realistic ISC within the Glauber approach tp FSI,  based on a linked cluster expansion \cite{ref:ciofi1} \cite{ref:cladahiko} are in progress and will be published elsewhere \cite{ref:ciofi2}.

\acknowledgements
We would like to thank N. N.  Nikolaev for useful comments and discussions.

 This work is in part supported by the Grant of Sapporo Gakuin University. H.M. thanks  INFN, Sezione di Perugia,  for the hospitality.

\begin{table}
 \caption{The results of the nuclear transparency T.}
 \label{tab:transparency}
  \begin{tabular}{ccccc}
   $T_0$ & $T_{ATMS}$ & $T_{Jastrow}$ & $T_S$ & $T_{ATMS}($no spect.$)$ \\
   \tableline
   0.754 & 0.778      & 0.806       & 0.780   & 0.778    
  \end{tabular}
\end{table}

\begin{table}
 \caption{Higer order contributio of Glauber multiple scattering series to the nuclear transparency T.}
 \label{tab:Glauber}
  \begin{tabular}{rcccc}
         & $\Delta G_1$ & $\Delta G_2$ & $\Delta G_3$ & Total \\
   \tableline
   1     & -0.2566     & 0.0363      & -0.0021     & 0.778   
  \end{tabular}
\end{table}

\begin{figure}[htbp]
	\caption{The realistic correlation function in the triplet $S$ state $^3\tilde{w}_s(r)=^{3}w_s(r)-(5/6)u(r)$ appearing in Eq.~(\ref{eq:atms_onshell})(full) compared with the Jastrow-type correlation function $f_J(r)$ defined by Eq.~(\ref{eq:jastrow}) (dashed). }
	\label{fig:correlation}
\end{figure}

\begin{figure}[htbp]
	\caption{The realistic momentum distribution (full) compared with the  momentum distribution obtained with the phenomenological Jastrow wave function (dashed).}
\label{fig:ennek}
\end{figure}

\begin{figure}[htbp]
	\caption{The parallel momentum distributions ($\theta=0^{\circ}$) corresponding to the harmonic oscillator model wave function (a) and to the realistic correlated variational wave function (Eq.~(\ref{eq:atmswf})) (b). The dashed curves represent the undistorted(No FSI) momentum distributions, whereas the full curves include FSI.}
\label{fig:gr+FSI}
\end{figure}

\begin{figure}[htbp]
	\caption{Comparison of the parallel ($\theta=0^{\circ}$), antiparallel 
($\theta=180^{\circ}$), and perpendicular ($\theta=90^{\circ}$) distorted momentum distributions  calculated using realistic (Realistic)((a) and (b)) and Jastrow (Jastrow)((c) and (d)) ground state wave functions including (FSI) and omitting (No FSI) final state interactions.}
	\label{fig:mdis_overview}
\end{figure}

\begin{figure}[htbp]
	\caption{The overall comparison between the distorted momentum distributions calculated using realistic (full) and Jastrow (dashed) ground state wave functions. Here (a)$\sim$(c) corresponds to the parallel($\theta=0^{\circ}$), perpendicular($\theta=90^{\circ}$) and antiparallel($\theta=180^{\circ}$) kinematics respectively.}
	\label{fig:comparison}
\end{figure}

\begin{figure}[htbp]
	\caption{The ratio between the distorted and undistorted momentum distributions $R({\bf k})=n_D({\bf k})/n({\bf k})$ corresponding to the realistic  (Realistic) and Jastrow (Jastrow) ground state wave functions (note the different  vertical scales).}
	\label{fig:ratio}
\end{figure}

\begin{figure}[htbp]
	\caption{The angular dependence of the distorted momentum distributions $n_D({\bf k})$ at fixed value of the missing momentum $ |{\bf k_m}|  \equiv k=3.0 fm^{-1}$. The various curves show the contributions from the different $^4$He waves. The practically constant value of the $D$ wave contribution demonstrates that FSI mostly act on the $S$ wave.}
	\label{fig:angular}
\end{figure}

\begin{figure}[htbp]
	\caption{The forward-backward asymmetry $A_{FB}(k)$ defined by Eq.~(\ref{eq:afb}) calculated with realistic (full) and Jastrow (dashed)  wave functions. The dot-dashed and short-dashed curves correspond to the realistic wave function with modified $D$-wave probability, namely $P_D=10.0\%$ and $15.0\%$ respectively.}
	\label{fig:asymmetry}
\end{figure}

\newpage

\begin{figure}
\begin{center}
   \epsfxsize=14cm \epsfbox{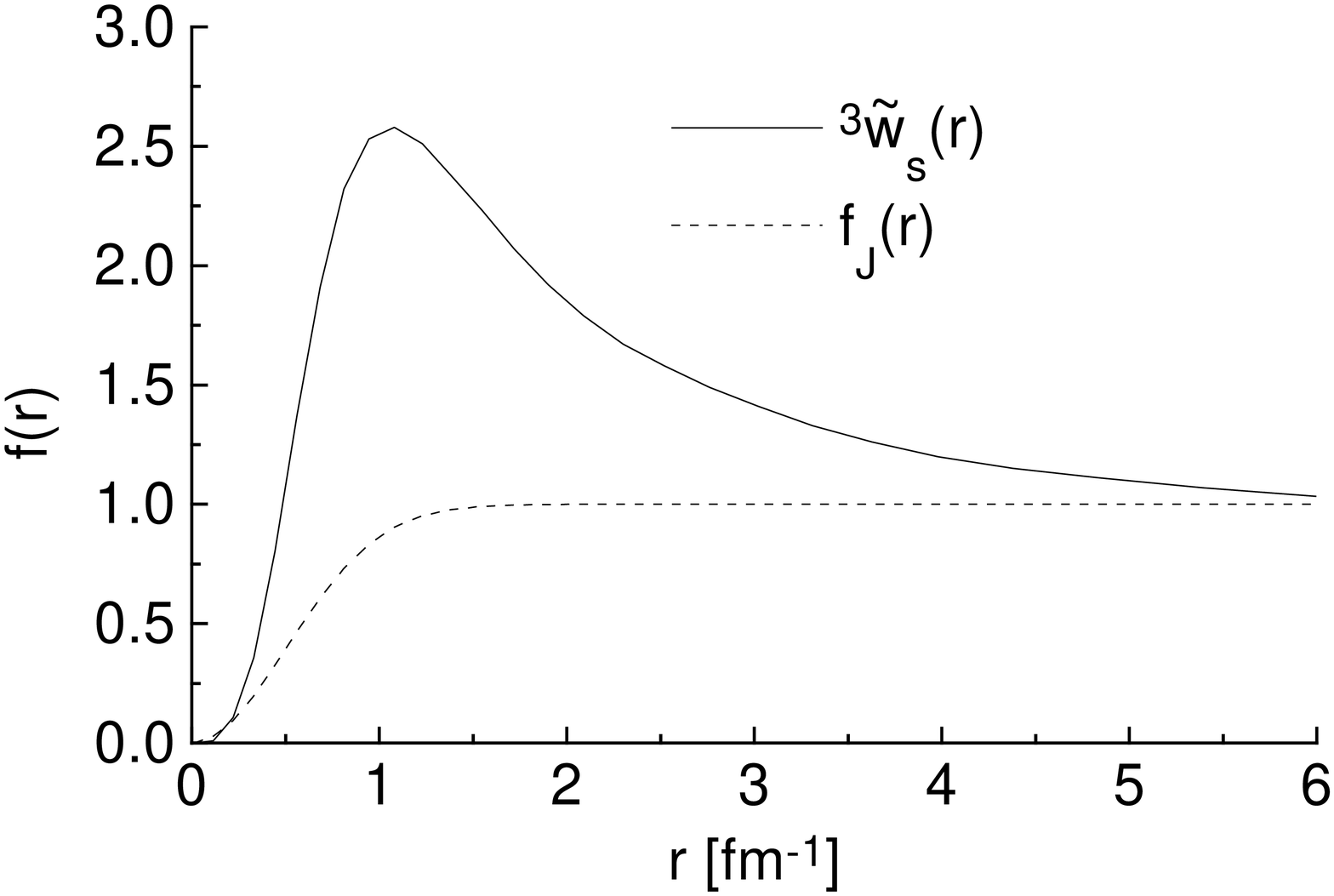}

   Fig. 1
\end{center} 
\end{figure}

\begin{figure}
\begin{center}
   \epsfxsize=14cm \epsfbox{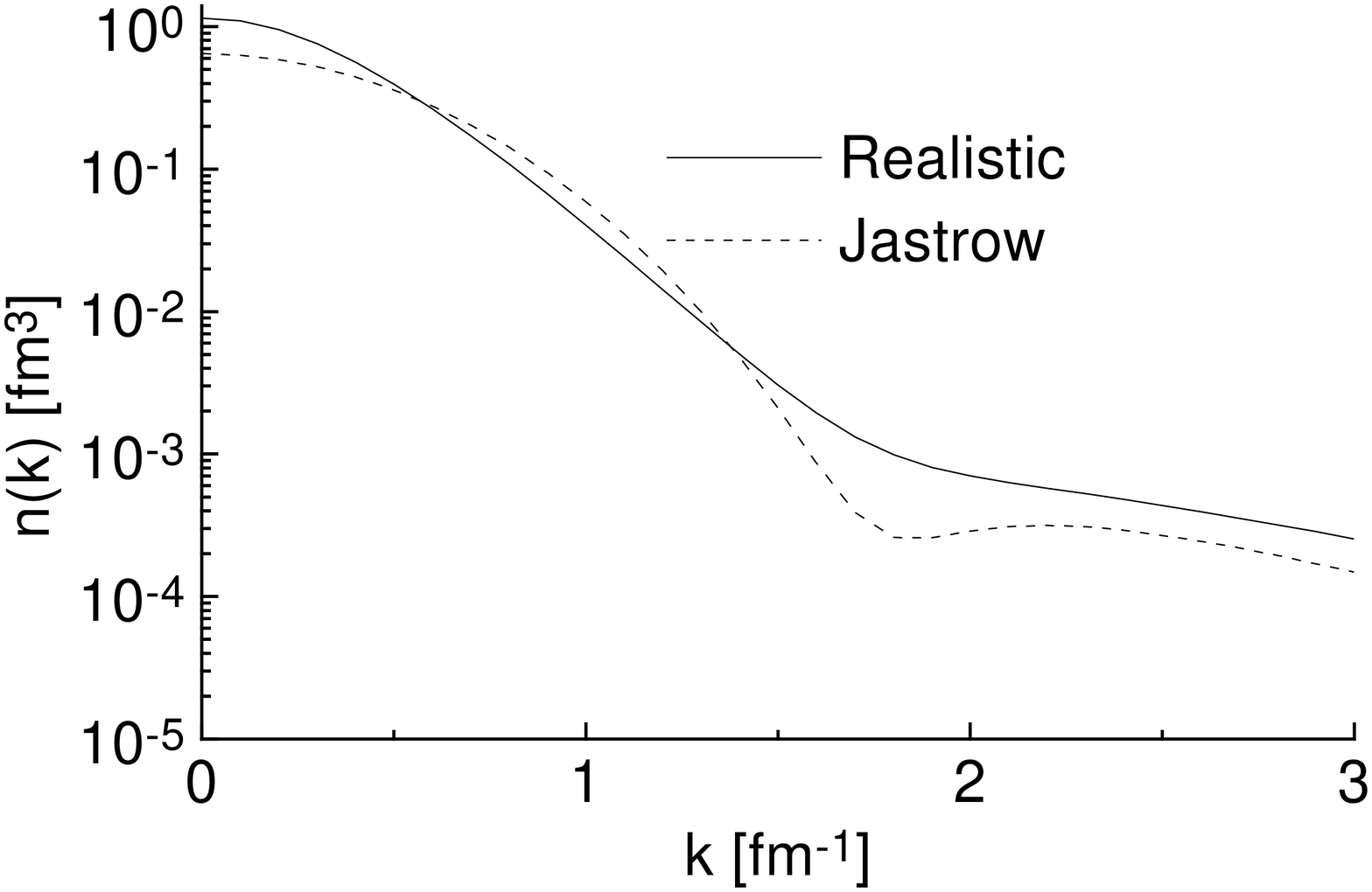}

 Fig. 2
\end{center} 
\end{figure}

\begin{figure}
\begin{center}
   \epsfxsize=15cm \epsfbox{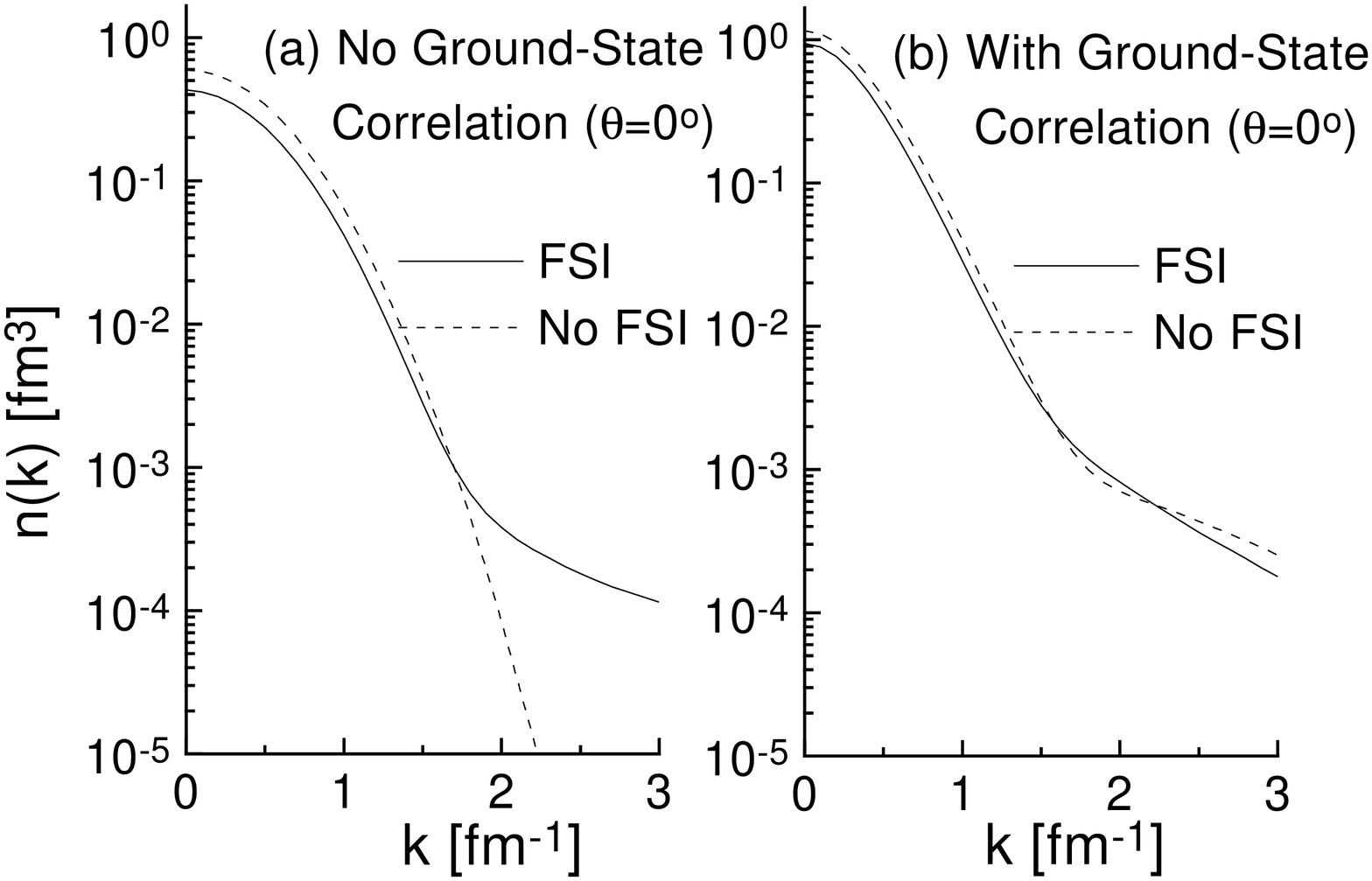}

   Fig.3
\end{center} 
\end{figure}

\begin{figure}
\begin{center}
   \epsfxsize=15cm \epsfbox{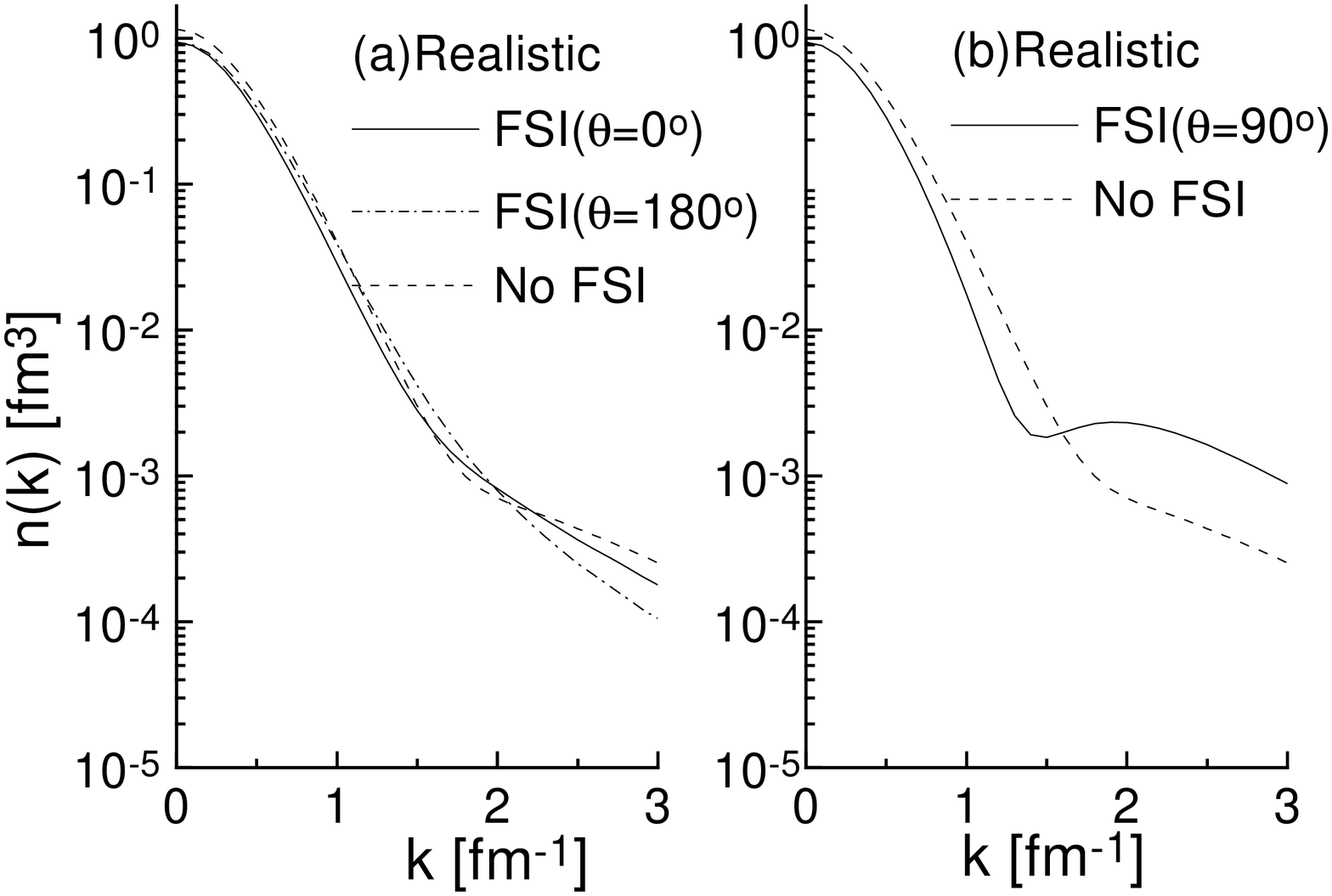}
   \epsfxsize=15cm \epsfbox{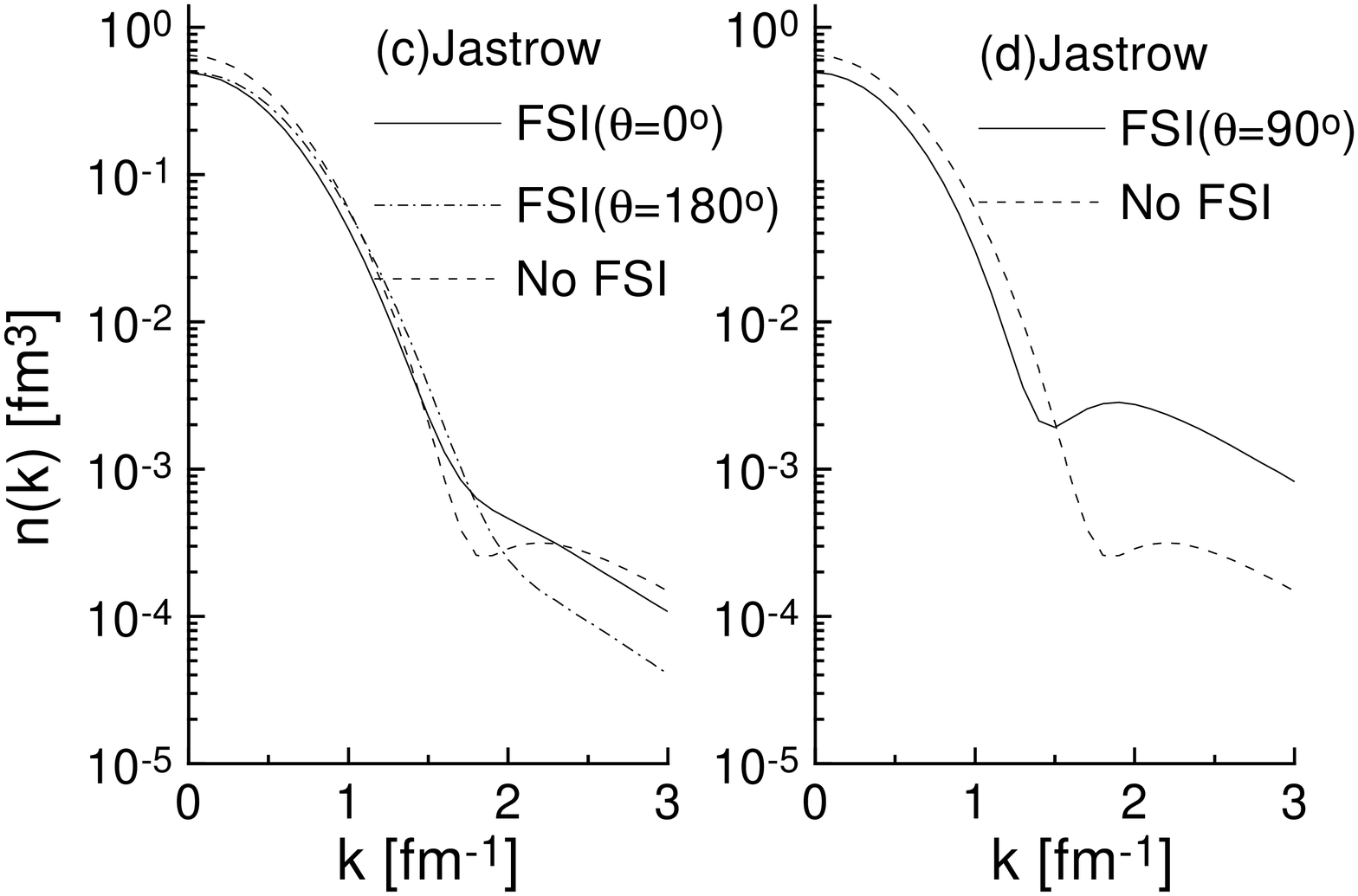}

   Fig. 4
\end{center} 
\end{figure}

\newpage
\begin{figure}
\begin{center}
   \epsfxsize=10cm \epsfbox{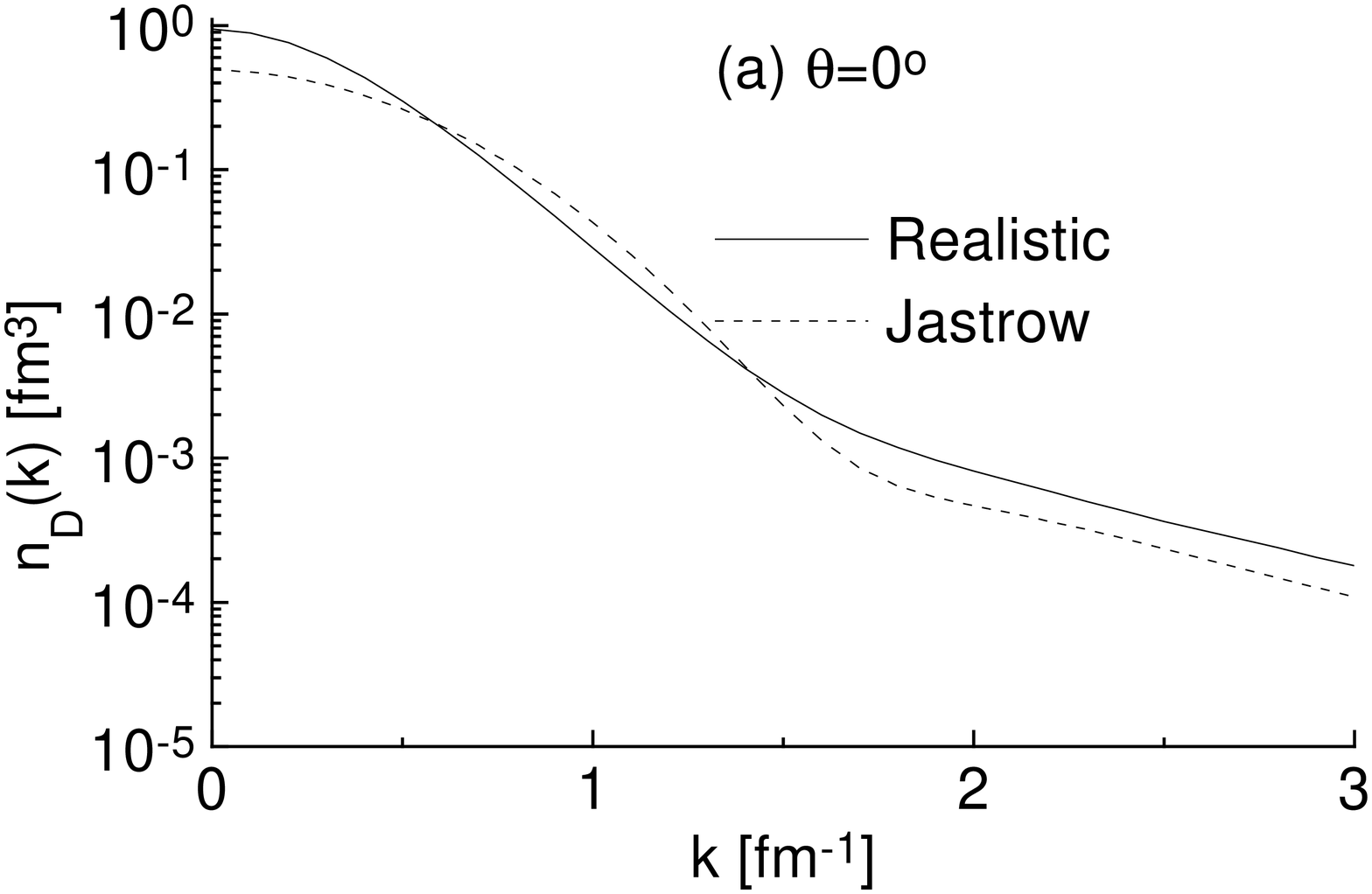}
   \epsfxsize=10cm \epsfbox{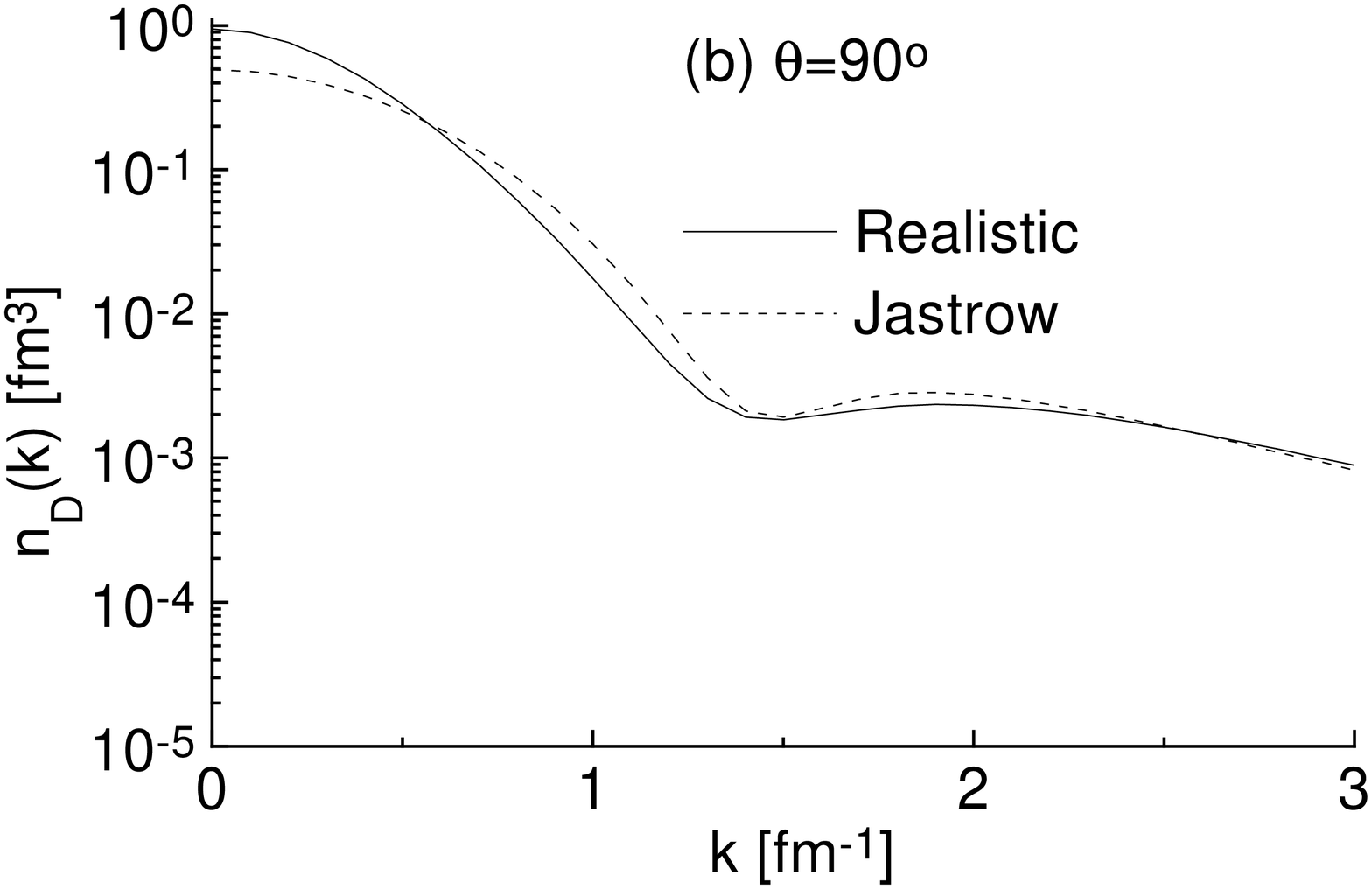}
   \epsfxsize=10cm \epsfbox{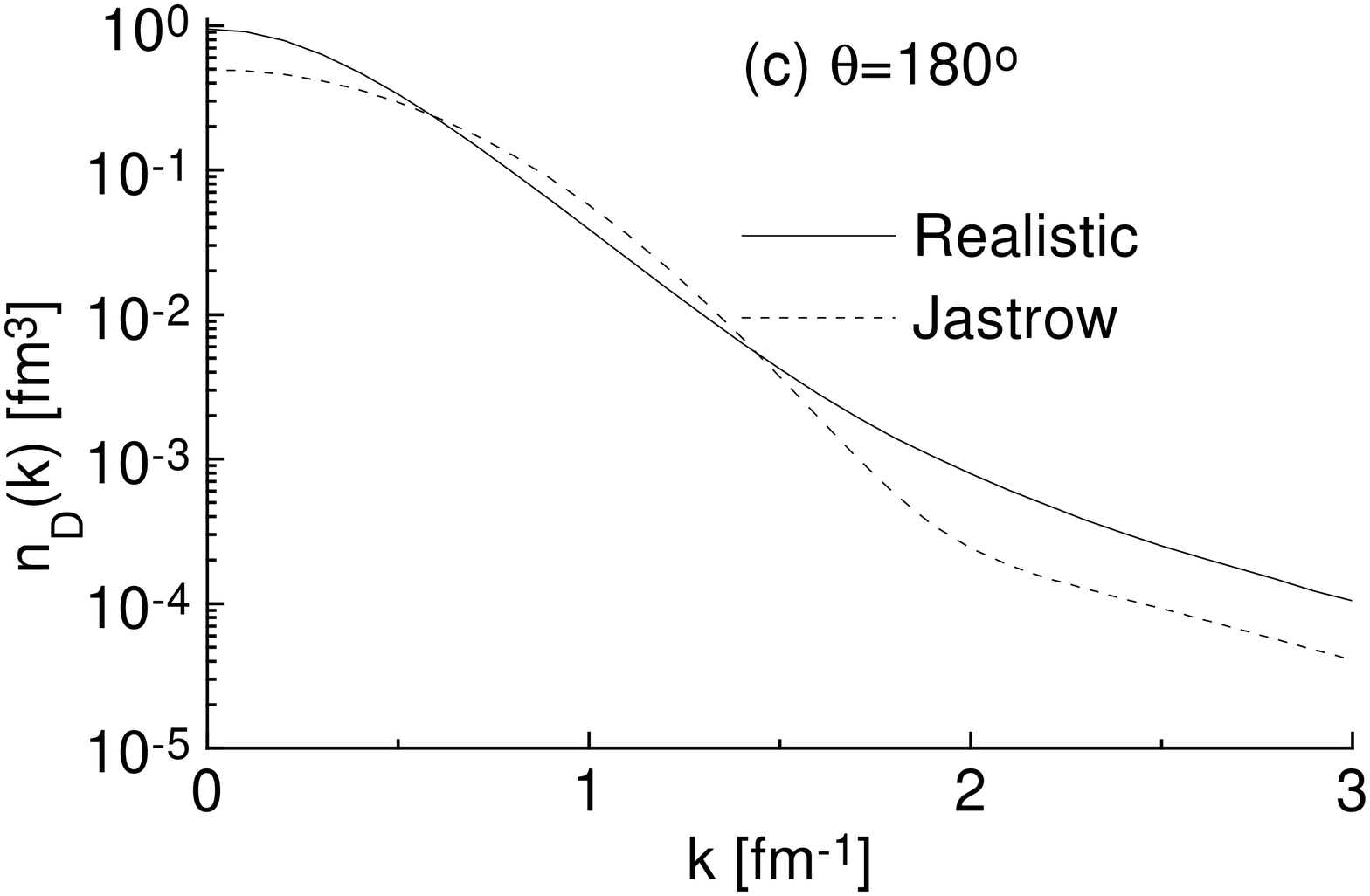}

  Fig. 5
\end{center} 
\end{figure}

\begin{figure}
\begin{center}
\epsfxsize=14cm \epsfbox{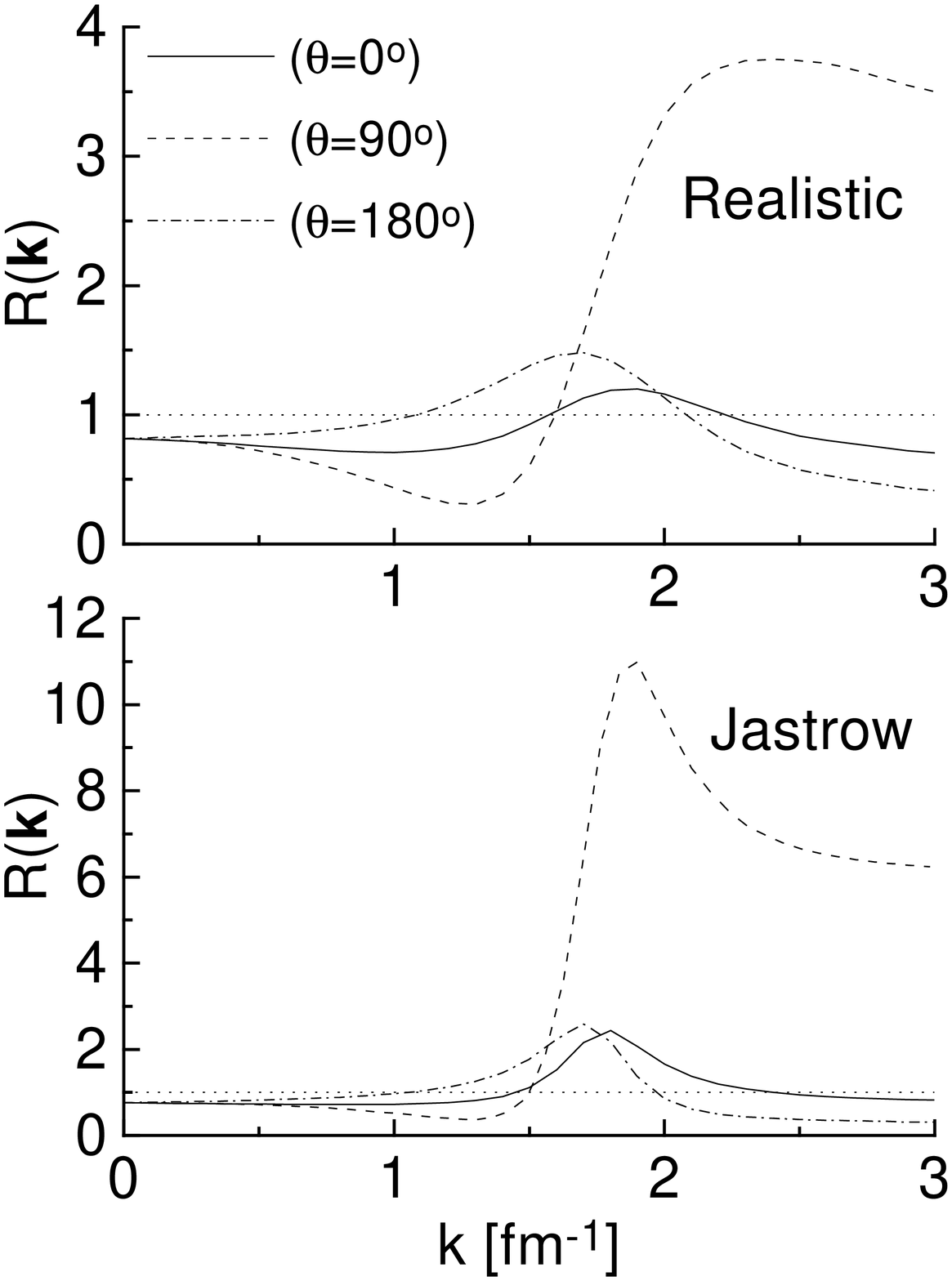}

Fig. 6
\end{center} 
\end{figure}
 
   \begin{figure}
\begin{center}
\epsfxsize=14cm \epsfbox{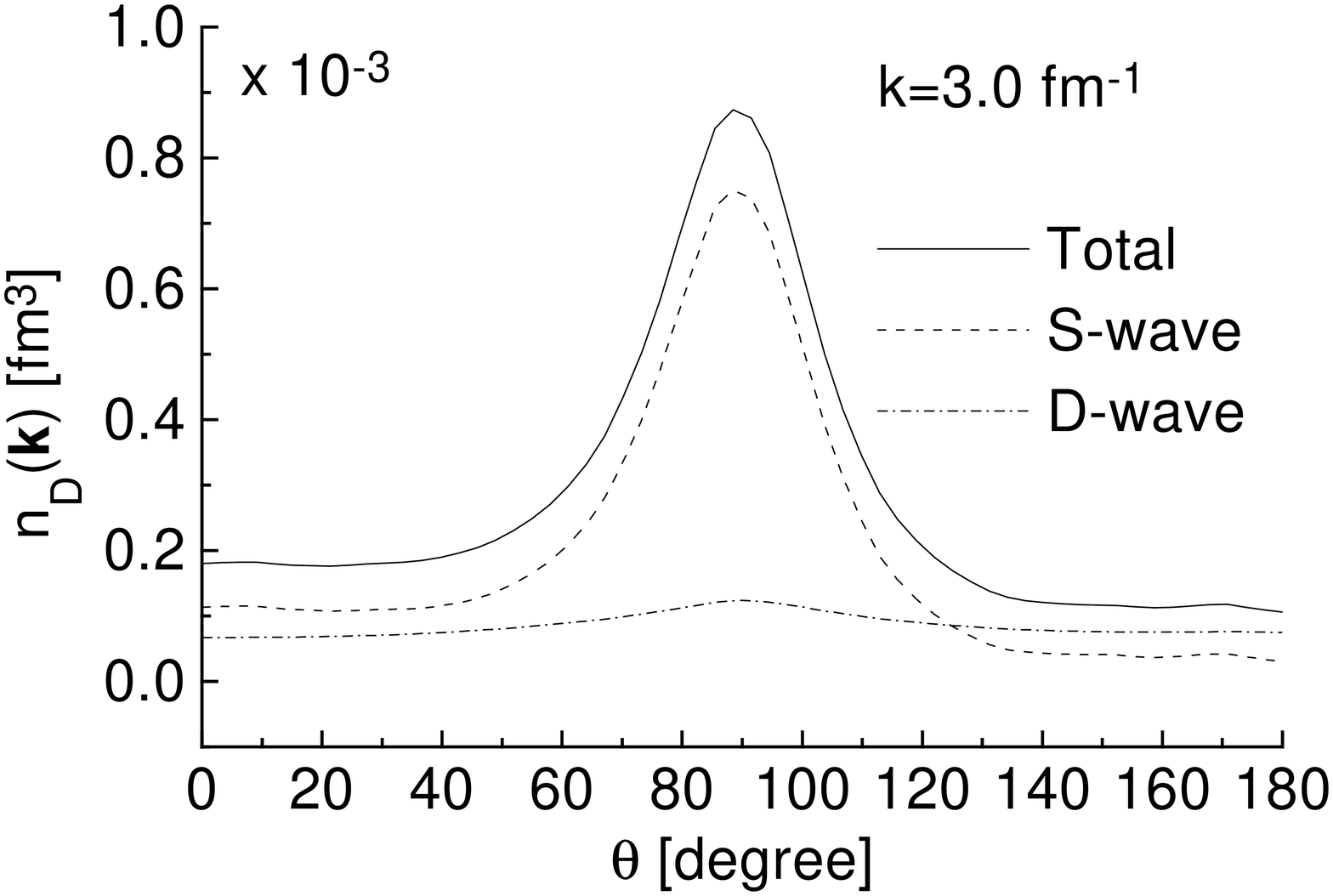}

Fig. 7  
\end{center} 
\end{figure}
 
\begin{figure}
\begin{center}
\epsfxsize=14cm \epsfbox{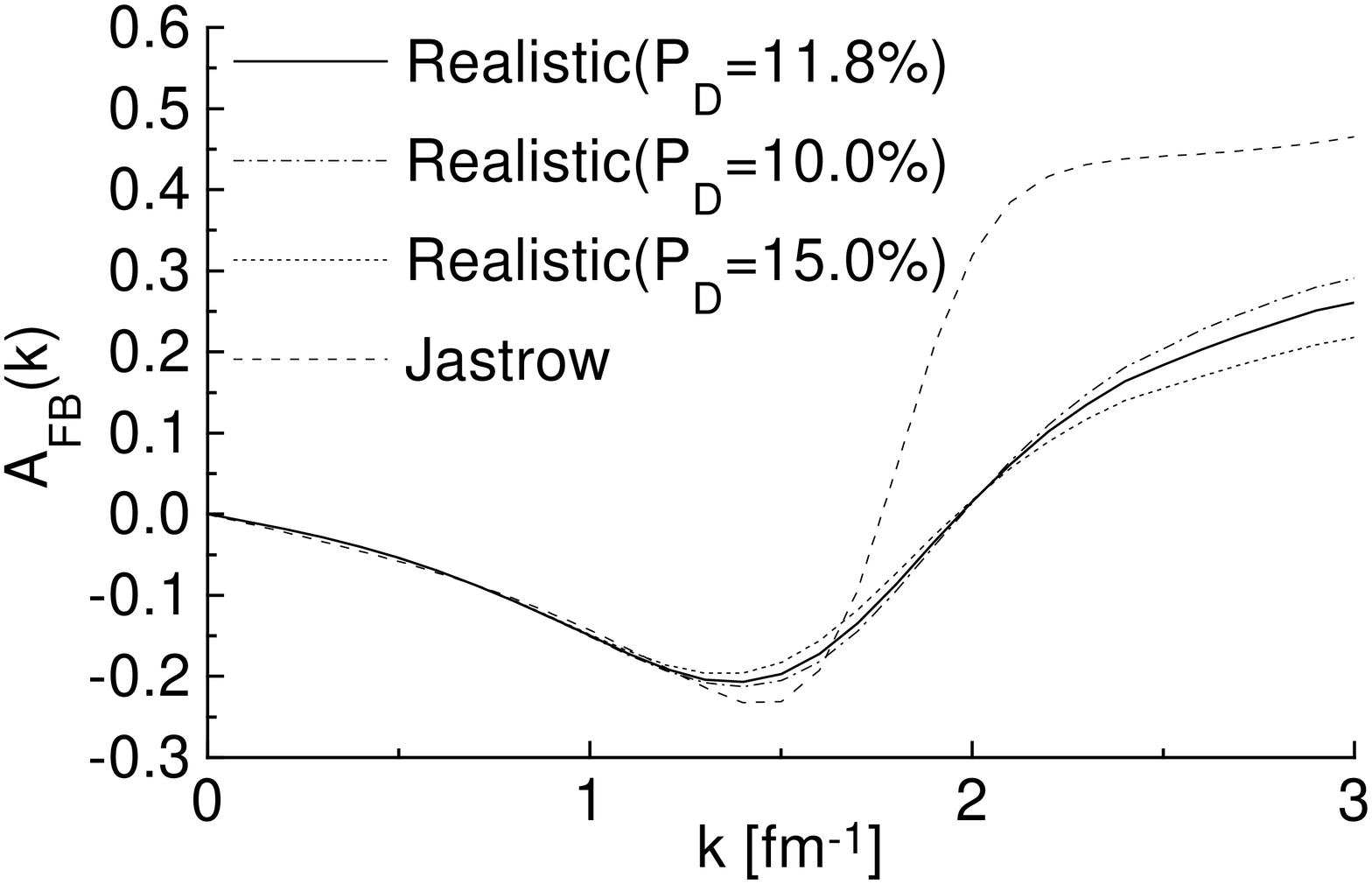}

  Fig. 8
\end{center} 
\end{figure}

\end{document}